\documentclass[conference]{IEEEtran}
\IEEEoverridecommandlockouts

\usepackage{array} 

\usepackage{graphicx} 
\usepackage{tabularx} 
\usepackage{booktabs}

\usepackage{cite}
\usepackage{amsmath,amssymb,amsfonts}
\usepackage{algorithmic}
\usepackage{graphicx}
\usepackage{textcomp}
\usepackage{xcolor}
\def\BibTeX{{\rm B\kern-.05em{\sc i\kern-.025em b}\kern-.08em
    T\kern-.1667em\lower.7ex\hbox{E}\kern-.125emX}}
\begin{document}

\title{SSM2Mel: State Space Model to Reconstruct Mel Spectrogram from the EEG\\

\thanks{
$^*$Corrresponding author

This work is supported by the {STI 2030—Major Projects (No. 2021ZD0201500)}, the National Natural Science Foundation of China (NSFC) (No.62201002, 6247077204), Excellent Youth Foundation of Anhui Scientific Committee (No. 2408085Y034), Distinguished Youth Foundation of Anhui Scientific Committee (No. 2208085J05), Special Fund for Key Program of Science and Technology of Anhui Province (No. 202203a07020008), Cloud Ginger XR-1.

}
}

\author{
\IEEEauthorblockN{\textit{Cunhang Fan$^{1}$, Sheng Zhang$^{1}$, Jingjing Zhang$^{1}$, Zexu Pan$^{2}$, Zhao Lv$^{1}$}}
\IEEEauthorblockA{$^{1}$School of Computer Science and Technology, Anhui University, Hefei, China\\
$^{2}$Alibaba group, Singapore\\
}
}

\maketitle

\begin{abstract}
Decoding speech from brain signals is a challenging research problem that holds significant importance for studying speech processing in the brain. Although breakthroughs have been made in reconstructing the mel spectrograms of audio stimuli perceived by subjects at the word or letter level using non-invasive electroencephalography (EEG), there is still a critical gap in precisely reconstructing continuous speech features, especially at the minute level. To address this issue, this paper proposes a State Space Model (SSM) to reconstruct the mel spectrogram of continuous speech from EEG, named SSM2Mel. This model introduces a novel Mamba module to effectively model the long sequence of EEG signals for imagined speech. In the SSM2Mel model, the S4-UNet structure is used to enhance the extraction of local features of EEG signals, and the Embedding Strength Modulator (ESM) module is used to incorporate subject-specific information. Experimental results show that our model achieves a Pearson correlation of 0.069 on the SparrKULee dataset, which is a 38\% improvement over the previous baseline.
\end{abstract}

\section{Introduction}
Speech is the foundation of human communication. Understanding how the brain processes speech is essential\cite{b1,b2,b31,b3}. Researchers have explored various strategies for decoding brain signals to extract semantic information, such as words, text, and natural language\cite{b4,b5,b6}. Among the array of brain signal recording technologies, invasive techniques offer detailed views of neural activity, but are limited by the necessity for surgical implantation\cite{b7}. In contrast, EEG, being a non-invasive technique, is safer and more readily accepted\cite{b8,b81}.

Decoding semantic information from EEG signals shows its potential in \cite{b9,b111,b112}, where researchers use a multi-channel convolutional neural network (MC-CNN) framework to identify the grammatical categories of imagined speech (verbs or nouns) from EEG signals. Following this, in \cite{b10}, researchers process and classify EEG signals of imagined numbers using multi-layer bidirectional recurrent neural networks (RNNs). These studies make significant progress in improving decoding accuracy, but they mostly focus on word or letter-level decoding\cite{b11}. Building on this foundation, researchers at \cite{b12} use a deep learning model integrated with multiple CNN blocks to attempt to recover the speech envelope heard by listeners at the minute level, marking an important step in the field of long-duration continuous imagined speech decoding. Subsequently, in \cite{b13}, researchers propose a transformer-based feedforward network to further improve the accuracy of speech envelope reconstruction. In the exploration of recovering speech features from EEG signals, \cite{b14} not only continues this research direction but also expands the scope of research, attempting to reconstruct more complex speech features using mel spectrograms. The ConvConcatNet model wins first place in the 2024 ICASSP Auditory Challenge, nevertheless the accuracy of reconstructing mel spectrograms from EEG signals still needs to be improved.

Addressing the challenges of decoding long-duration imagined speech and improving the accuracy of mel spectrogram recovery, this paper proposes a novel model called SSM2Mel, which is based on the State Space Model to reconstruct mel spectrogram from EEG. To address the issue of long sequence modeling over extended periods, we utilize the Selective State
Space Models (Mamba) module\cite{b15}, which excels at handling long sequence data with high efficiency, and combine it with Multi-Head Self-Attention (MHSA)\cite{b16}.  Furthermore, in order to capture local features and short-term dependencies, enabling the model to more accurately locate and reconstruct brain activity patterns related to speech production, we propose for the first time a U-Net structure\cite{b17} based on Structured State Space Sequence (S4) layers \cite{b18} for the processing of EEG signals. To capture correlations between different samples and enhance the model's sensitivity to individual differences, we also introduce an external attention mechanism\cite{b19} and an ESM module\cite{b20}.

\begin{figure*}
    \centering
    \includegraphics[width=0.9\textwidth]{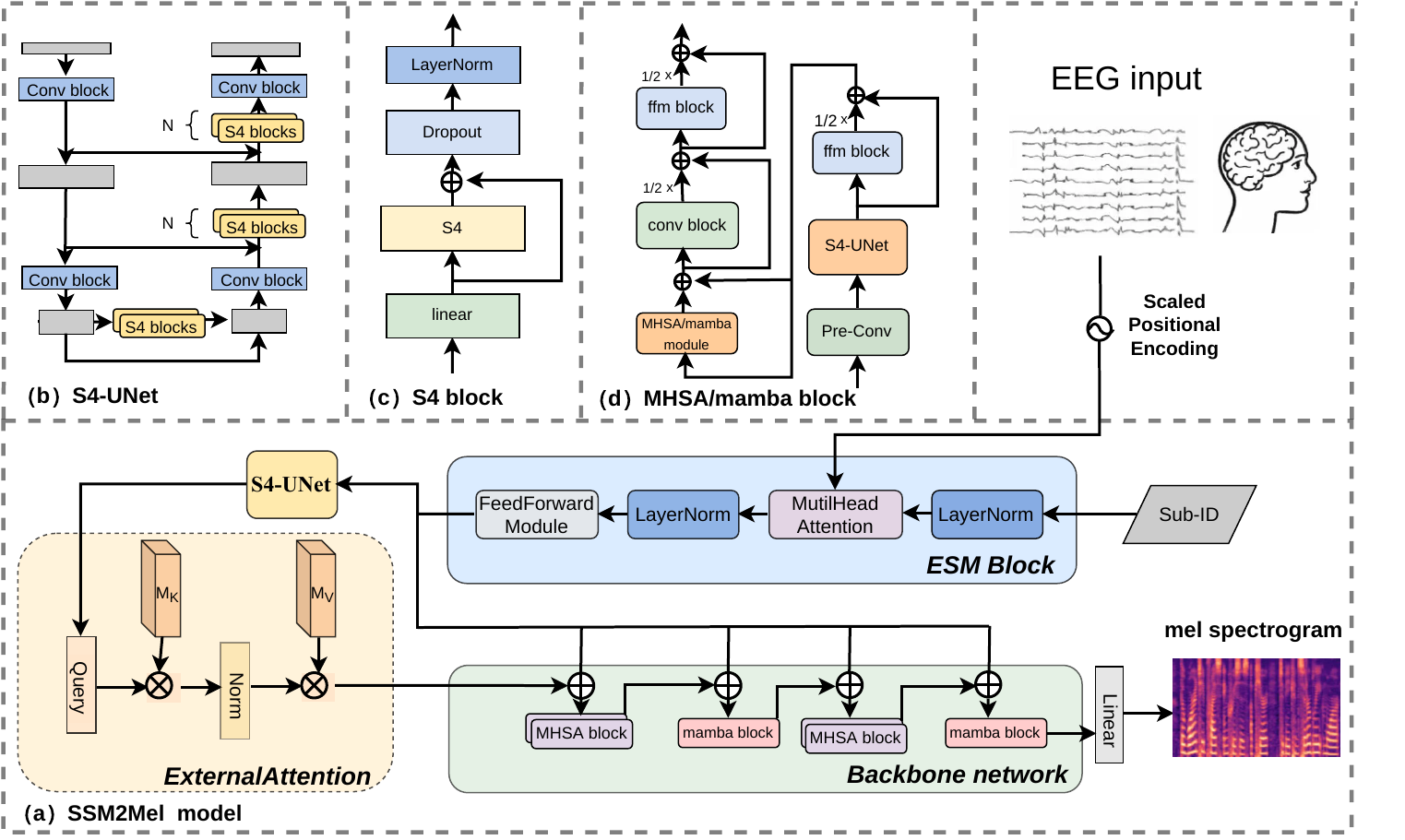}
    \caption{The overall framework of this study. This framework is composed of an ESM block, an S4-UNet, ExternalAttention, and a backbone network. In which $\otimes$ represents matrix multiplication, and $\oplus$ represents residual connection.}
    \label{fig:enter-label}
\end{figure*}

The contributions of this work are threefold:

\begin{itemize}
  \item  We innovatively use the Mamba model to efficiently model long sequences of EEG signals, exploring the best practices associated with it.
  \item We innovatively apply the S4-based U-Net network structure to EEG signal processing to capture local features and the short-term dependencies of these features.

  \item Our experiments on the ICASSP 2024 Auditory EEG Challenge dataset demonstrate that our model's performance significantly outperforms the current state-of-the-art models.
\end{itemize}

\section{Methods}
In this section, we introduce the SSM2Mel model, an advanced neural network framework, designed to convert EEG signals into mel spectrograms, as shown in Fig.1. The model consists of a series of interconnected modules, each with a specific function in the feature extraction and generation process. Here, we provide a detailed description of the model's architecture and its components.

\subsection{Overview of SSM2Mel}

As shown in Fig.1(a), the model we proposed consists of three main parts. The first part is the ESM module, which receives EEG signals and sub-IDs as inputs, with sub-IDs being the unique identifiers for each subject in the dataset. The ESM module enhances the personalization and adaptability of features through dynamic adjustments, providing input features that include physiological and semantic information for subsequent tasks. The EEG signals processed by the ESM are then fed into the second part of the model, where the S4-UNet and ExternalAttention modules perform preliminary feature extraction. The S4-UNet captures local features of EEG signals with its structure, while the ExternalAttention module increases sensitivity to individual differences through its external attention mechanism, making the extracted features rich in spatiotemporal information and sensitive to individual differences. The third part of the model is a backbone network composed of Mamba and MHSA modules. We innovatively use the Mamba module and integrate it with MHSA, forming an efficient processing procedure. This network receives pre-extracted features and integrates raw features processed solely by the ESM module to learn a broader range of data representations, and improves training stability through diverse inputs. Finally, the output of the backbone network is summarized into a single mel spectrogram through a linear layer.

\subsection{Embedding Strength Modulator}
EEG signals exhibit significant individual variability, with brain activities and responses to external stimuli varying across subjects\cite{b21}. By incorporating each subject's unique ID as part of the model input, we enhance the model's ability to capture these individual differences, thereby improving the accuracy of imagined speech reconstruction\cite{b13}. Drawing inspiration from the ESM module in bilingual TTS systems, which adjusts embedding strength to precisely capture subtle differences in bilingual speech\cite{b20}, we adopt this technology to integrate specific subject information and customize model responses based on individual subject characteristics. The ESM module's function of modulating voice and intonation inspires us to explore its potential to adapt to different subject brain activity patterns, enabling our model to dynamically optimize its responses to individual differences. The specific structure of the ESM module is shown in Fig.1(a).
Formally, starting from the initial EEG input \( P_0 \) with scaled positional encoding, and using an embedding \( SE \) based on the subject's identity with one-hot encoding, the calculations for the first sub-network \( M_0 \) and the second sub-network \( F_0 \) are as follows:
    
\begin{equation}
M_{o} = MH\left(P_o, LN\left(SE\right) + SE\right)
\end{equation}
\begin{equation}
F_{o} = FFN\left(LN\left(M_{o}\right)\right) + M_{o}
\end{equation}
where \( MH(query, key, value) \), \( FFN(\cdot)\) and \( LN(\cdot) \) are multi-head attention, feedforward network and layer normalization, respectively.

\subsection{S4-UNet Block}
\cite{b22} introduces a structure composed of repeated S4 blocks for speech enhancement. Inspired by this, we propose the S4-UNet model, as shown in Fig.1(b). This structure involves injecting S4 layers into the U-Net architecture. The U-Net has multiple down-sampling and up-sampling layers, processing data through n S4 blocks before up-sampling. The specific construction of the S4 block is shown in Fig.1(c). The core of the S4 model is its ability to combine the advantages of CNNs and RNNs. In the S4-UNet, S4 blocks process signals within local neighborhoods, which is crucial for capturing local features and short-term dependencies. Consequently, the S4-UNet plays a significant role in enhancing the convolutional modules and pre-extraction capabilities of in our model.

In addition to this, we also apply the ExternalAttention after the pre-extracted S4-UNet. This mechanism effectively encodes signals through key (\( Mk \) ) and value (\( Mv \)) storage units while maintaining linear complexity. It acts as a preprocessing module, helping the network to capture individual-specific features early on, enhancing sensitivity to individual differences in EEG signals, and maintaining computational efficiency.

\subsection{Backbone Network}
Inspired by the macaron-like structure of the Conformer model\cite{b23}, we develop the MHSA/Mamba block, as shown in Fig.1(d). Similar to the Conformer, our model integrates core components such as feedforward mechanisms, multi-head attention, and convolutions. Notably, our Mamba block replaces the traditional multi-head attention module with the Mamba module, which requires additional nonlinearity to learn advanced semantic information\cite{b24} and is best used as an alternative to MHSA. Although the Conformer already captures long-range dependencies, its performance can be further enhanced through structural improvements\cite{b25}. Our model introduces pre-convolution and S4-UNet before the Conformer architecture to strengthen the convolutional module, making it more sensitive to local features in addition to its own convolutional capabilities.

The S4 model and the Mamba model are architectures based on the SSM for handling sequential data. The SSM model captures the dynamics of sequence data with the following state space representation:
\begin{equation}
h_{t} = Ah_{t-1}  +Bx_{t} 
\end{equation}
\begin{equation}
y_{t} = Ch_{t} + D x_{t} 
\end{equation}
here, \( h_t \) is the hidden state, \( x_t \) is the input, and \( y_t \) is the output at time step \( t \). The matrices \( A \), \( B \), \( C \), and \( D \) define the model's dynamics. Based on the SSM model, the S4 model introduces a structured state space to improve the processing of sequential data, with its core formula represented as:
\begin{equation}
h_t = \bar{A} h_{t-1} + \bar{B} x_t 
\end{equation}
\begin{equation}
y_{t} = Ch_{t} 
\end{equation}
in this formula, \( \bar{A} \) and \( \bar{B} \) are the discrete-time matrices derived from the continuous-time matrices \( A \) and \( B \). while \( h_t \), \( x_t \) and C retain the same definitions as in the SSM model. Compared to the S4 model, the Mamba model introduces parameterized matrices, allowing the \( A \), \( B \), and \( C \) matrices to vary based on the input.

New hybrid models such as Jamba\cite{b26} and Zamba\cite{b27} add attention layers to enhance performance. Experiments show that combining Mamba layers with attention layers outperforms using them individually, indicating that attention mechanisms and  SSMs complement each other. Therefore, we design a backbone network that integrates both MHSA and Mamba modules.

\subsection{Loss Function}
For the loss function, in order to ensure stable training, we combine the \( \mathcal{L}_{1} \) loss with the negative Pearson correlation loss for model training. The Pearson correlation coefficient loss measures the correlation between the predicted and actual mel spectrogram values of the stimulus audio, while the \( \mathcal{L}_{1} \) loss measures the absolute error between predicted and true values. The overall training loss L is defined as follows:
\begin{equation}
\mathcal{L} = -R+\alpha \times \mathcal{L}_{1} 
\end{equation}
\begin{equation}
\mathcal{L}_{1} = \frac{1}{n}  {\textstyle \sum_{i=1}^{n}} \left | \hat{y}_{i} -y_{i}   \right | 
\end{equation}
where $\hat{y}_{i}$ represents the predicted value of the mel spectrogram, ${y}_{i}$ represents the actual value, $R$ represents the Pearson correlation coefficient, and $\alpha$ is a hyperparameter that balances the weights of the two parts of the loss function.

\section{Experiments and Results}
\subsection{Dataset}
We utilize the dataset from the ICASSP 2024 Auditory EEG Challenge, SparrKULee\cite{b28}, which includes EEG data from 85 Dutch-speaking participants with normal hearing. Each subject undergoes 8 to 10 trials lasting approximately 15 minutes, listening to stories told in fluent Flemish (Belgian Dutch). We apply multichannel Wiener filtering to the raw data to remove artifacts, then re-reference the EEG signals to a common average, and downsample the EEG signals to 64 Hz. The dataset is mainly divided into three parts: training (80\%), validation (10\%), and testing (10\%), to optimize model training and evaluate performance.

\begin{table}[h]
\centering
\caption{Peason Correlation Values of Mel Spectrogram Recovery} 
\resizebox{0.5\textwidth}{!}{
\addtolength{\tabcolsep}{-5pt}
\begin{tabular}{@{}ccc@{}}
\toprule
Model                   & Pearson correlation & \begin{tabular}[c]{@{}c@{}}Performance\\ improvement (\%)\end{tabular} \\ \midrule
VLAAI\cite{b12}                   & 0.050               & -                                                                   \\
HappyQuokka\cite{b13}              & 0.052               & 4                                                               \\
ConvConcatNet\cite{b14}            & 0.058               & 16                                                                    \\
ConvConcatNet ensembled\cite{b14}  & 0.063               & 26                                                                   \\ \midrule
SSM2Mel (proposed)               & \textbf{0.069}               & \textbf{38}                                                                      \\ \bottomrule
\end{tabular}}
\end{table}

\begin{table}[h]
\centering
\caption{Performance Comparison in Speech Envelope Recovery} 
\resizebox{0.5\textwidth}{!}{

\addtolength{\tabcolsep}{1.5pt}
\begin{tabular}{@{}ccc@{}}
\toprule
Model       & Pearson correlation & \begin{tabular}[c]{@{}c@{}}Performance\\ improvement (\%)\end{tabular} \\ \midrule
VLAAI        & 0.161               & -                                                                  \\
HappyQuokka & 0.185               & 14.9                                                                  \\ \midrule
SSM2Mel (proposed)    & \textbf{0.208}               & \textbf{29.2}                                                                      \\ \bottomrule
\end{tabular}}
\end{table}

\subsection{Implementation Details}
We implement our proposed model using PyTorch. We train our model for 1000 epochs using the Adam optimizer with an initial learning rate of 0.0005. We employ the StepLR scheduler, which automatically decreases the learning rate by a factor of 0.9 every 50 epochs. During the training process, we utilize 5-second signal segments, randomly cropped from each EEG/speech mel spectrogram segment, to ensure stable training. For inference, we divide the input signal into several 5-second segments, process each segment individually, and then concatenate the outputs to form the complete mel spectrogram.

\subsection{Comparison with Baselines}

Table I summarizes the results of the baseline and our proposed model in recovering the mel spectrogram of speech, comparing them with those of three state-of-the-art models: VLAAI\cite{b12}, HappyQuokka\cite{b13}, and ConvConcatNet\cite{b14}. Our model performs exceptionally well in the reconstruction task, achieving a Pearson correlation coefficient of 0.068. Compared to VLAAI, HappyQuokka, and ConvConcatNet, our method shows an increase in the Pearson correlation coefficient by 0.019, 0.017, and 0.011, respectively, further demonstrating the effectiveness of our approach. Although ConvConcatNet has achieved commendable results with the aid of ensemble learning, our method still manages to achieve higher correlation, which may be attributed to the unique design and optimization strategies of our model.It is worth noting that the "Performance improvement" metric in all our tables represents the percentage increase in performance relative to the VLAAI baseline model.

To further demonstrate the superiority of our proposed model, we compare our model with VLAAI and HappyQuokka on the same dataset for the task of recovering the subject's speech envelope, as shown in Table II. The results indicate that our proposed model still achieves the best results.

\subsection{Ablation Study}

We conduct ablation studies to verify the necessity of each module in our model, as shown in Table III. Removing the S4-UNet, ESM, or ExternalAttention modules all lead to a significant drop in the Pearson correlation coefficient, demonstrating that these modules are crucial for enhancing model performance. 

To further demonstrate the advantage of our approach that integrates multi-head attention modules with Mamba modules, we compare the performance of models using only the Mamba module, the MHSA module, and the combination of both, as shown in Table IV. The results indicate that using both types of modules significantly improves model performance, likely because the Mamba module strengthens the processing of temporal signal characteristics, while the MHSA module enhances the capture of contextual information. Our experimental results support the rationality of the current module combination and suggest that this design helps the model better handle EEG signals. Future research will further explore these preliminary findings.





\begin{table}[]
\centering
\caption{Ablation Study on the Impact of Network Modules} 
\resizebox{0.5\textwidth}{!}{

\addtolength{\tabcolsep}{1.3pt}

\begin{tabular}{@{}ccc@{}}
\toprule
Model                 & \multicolumn{1}{l}{Pearson correlation} & \multicolumn{1}{l}{\begin{tabular}[c]{@{}c@{}}Performance\\ improvement (\%)\end{tabular}} \\ \midrule
VLAAI                 & 0.050                                   & -                                                                                          \\ \midrule
w/o S4-UNet           & 0.057                                   & 14                                                                                         \\
w/o ExternalAttention & 0.061                                   & 22                                                                                         \\
w/o ESM               & 0.063                                   & 26                                                                                         \\
SSM2Mel (proposed)    & \textbf{0.069}                          & \textbf{38}                                                                                \\ \bottomrule
\end{tabular}}
\end{table}

\begin{table}[]
\centering
\caption{Comparative Performance of Integrated Modules} 
\resizebox{0.5\textwidth}{!}{

\addtolength{\tabcolsep}{1pt}

\begin{tabular}{@{}ccc@{}}
\toprule
Model           & \multicolumn{1}{l}{Pearson correlation} & \multicolumn{1}{l}{\begin{tabular}[c]{@{}c@{}}Performance\\ improvement (\%)\end{tabular}} \\ \midrule
VLAAI                 & 0.050           & -    \\  \midrule
all-mamba model & 0.051                                   & 2                                                                                       \\
all-MHSA model  & 0.057                                   & 14                                                                                         \\   
 SSM2Mel (proposed)         & \textbf{0.069}                                   & \textbf{38}                                                                                          \\ \bottomrule
\end{tabular}}
\end{table}

\section{Conclusion}

In this study, we propose the SSM2Mel framework for reconstructing mel spectrograms from long-duration EEG signals. Utilizing the superior performance of state space models in sequence modeling, the model achieves state-of-the-art performance on public datasets, with ablation experiments further validating the effectiveness of its strategies and components. This research provides a new framework for long-sequence decoding in cognitive science, neuroscience, and brain-computer interface fields, and offers a reference for future model design.


\begin{thebibliography}{00}

\bibitem{b1}
J.~G. Makin, D.~A. Moses, and E.~F. Chang, ``Machine translation of cortical activity to text with an encoder--decoder framework,'' \emph{Nature neuroscience}, vol.~23, no.~4, pp. 575--582, 2020.

\bibitem{b2}
J.~Tang, A.~LeBel, S.~Jain, and A.~G. Huth, ``Semantic reconstruction of continuous language from non-invasive brain recordings,'' \emph{Nature Neuroscience}, vol.~26, no.~5, pp. 858--866, 2023.

\bibitem{b31}
C.~Fan, J.~Zhang, H.~Zhang, W.~Xiang, J.~Tao, X.~Li, J.~Yi, D.~Sui, and Z.~Lv, ``Msfnet: Multi-scale fusion network for brain-controlled speaker extraction,'' in \emph{Proceedings of the 32nd ACM International Conference on Multimedia}, 2024, pp. 1652--1661.

\bibitem{b3}
B.~Wang, X.~Xu, L.~Zhang, B.~Xiao, X.~Wu, and J.~Chen, ``Semantic reconstruction of continuous language from meg signals,'' in \emph{ICASSP 2024-2024 IEEE International Conference on Acoustics, Speech and Signal Processing (ICASSP)}.\hskip 1em plus 0.5em minus 0.4em\relax IEEE, 2024, pp. 2190--2194.

\bibitem{b4}
N.~Xi, S.~Zhao, H.~Wang, C.~Liu, B.~Qin, and T.~Liu, ``Unicorn: Unified cognitive signal reconstruction bridging cognitive signals and human language,'' in \emph{Proceedings of the 61st Annual Meeting of the Association for Computational Linguistics (Volume 1: Long Papers)}, 2023, pp. 13\,277--13\,291.

\bibitem{b5}
A.~D{\'e}fossez, C.~Caucheteux, J.~Rapin, O.~Kabeli, and J.-R. King, ``Decoding speech perception from non-invasive brain recordings,'' \emph{Nature Machine Intelligence}, vol.~5, no.~10, pp. 1097--1107, 2023.

\bibitem{b6}
Y.~Duan, J.~Zhou, Z.~Wang, Y.-K. Wang, and C.-T. Lin, ``Dewave: discrete eeg waves encoding for brain dynamics to text translation,'' in \emph{Proceedings of the 37th International Conference on Neural Information Processing Systems}, 2023, pp. 9907--9918.

\bibitem{b7}
J.~R. Wolpaw, N.~Birbaumer, D.~J. McFarland, G.~Pfurtscheller, and T.~M. Vaughan, ``Brain--computer interfaces for communication and control,'' \emph{Clinical neurophysiology}, vol. 113, no.~6, pp. 767--791, 2002.

\bibitem{b8}
S.~Luo, Q.~Rabbani, and N.~E. Crone, ``Brain-computer interface: applications to speech decoding and synthesis to augment communication,'' \emph{Neurotherapeutics}, vol.~19, no.~1, pp. 263--273, 2023.

\bibitem{b81}
C.~Fan, H.~Zhang, W.~Huang, J.~Xue, J.~Tao, J.~Yi, Z.~Lv, and X.~Wu, ``Dgsd: Dynamical graph self-distillation for eeg-based auditory spatial attention detection,'' \emph{Neural Networks}, vol. 179, p. 106580, 2024.

\bibitem{b9}
S.~Datta and N.~V. Boulgouris, ``Recognition of grammatical class of imagined words from eeg signals using convolutional neural network,'' \emph{Neurocomputing}, vol. 465, pp. 301--309, 2021.

\bibitem{b111}
Q.~Ni, H.~Zhang, C.~Fan, S.~Pei, C.~Zhou, and Z.~Lv, ``Dbpnet: Dual-branch parallel network with temporal-frequency fusion for auditory attention detection,'' in \emph{Proceedings of the International Joint Conference on Artificial Intelligence (IJCAI 2024)}, 2024.

\bibitem{b112}
S.~Yan, H.~Zhang, X.~Yang, J.~Tao, Z.~Lv \emph{et~al.}, ``Darnet: Dual attention refinement network with spatiotemporal construction for auditory attention detection,'' \emph{arXiv preprint arXiv:2410.11181}, 2024.



\bibitem{b10}
N.~C. Mahapatra and P.~Bhuyan, ``Eeg-based classification of imagined digits using a recurrent neural network,'' \emph{Journal of Neural Engineering}, vol.~20, no.~2, p. 026040, 2023.

\bibitem{b11}
A.~Kamble, P.~H. Ghare, V.~Kumar, A.~Kothari, and A.~G. Keskar, ``Spectral analysis of eeg signals for automatic imagined speech recognition,'' \emph{IEEE Transactions on Instrumentation and Measurement}, vol.~72, pp. 1--9, 2023.



\bibitem{b12}
B.~Accou, J.~Vanthornhout, H.~V. hamme, and T.~Francart, ``Decoding of the speech envelope from eeg using the vlaai deep neural network,'' \emph{Scientific Reports}, vol.~13, no.~1, p. 812, 2023.

\bibitem{b13}
Z.~Piao, M.~Kim, H.~Yoon, and H.-G. Kang, ``Happyquokka system for icassp 2023 auditory eeg challenge,'' in \emph{ICASSP 2023-2023 IEEE International Conference on Acoustics, Speech and Signal Processing (ICASSP)}.\hskip 1em plus 0.5em minus 0.4em\relax IEEE, 2023, pp. 1--2.

\bibitem{b14}
X.~Xu, B.~Wang, Y.~Yan, H.~Zhu, Z.~Zhang, X.~Wu, and J.~Chen, ``Convconcatnet: a deep convolutional neural network to reconstruct mel spectrogram from the eeg,'' \emph{arXiv preprint arXiv:2401.04965}, 2024.

\bibitem{b15}
A.~Gu and T.~Dao, ``Mamba: Linear-time sequence modeling with selective state spaces,'' \emph{arXiv preprint arXiv:2312.00752}, 2023.



\bibitem{b16}
V.~Ashish, ``Attention is all you need,'' \emph{Advances in neural information processing systems}, vol.~30, p.~I, 2017.



\bibitem{b17}
O.~Ronneberger, P.~Fischer, and T.~Brox, ``U-net: Convolutional networks for biomedical image segmentation,'' in \emph{Medical image computing and computer-assisted intervention--MICCAI 2015: 18th international conference, Munich, Germany, October 5-9, 2015, proceedings, part III 18}.\hskip 1em plus 0.5em minus 0.4em\relax Springer, 2015, pp. 234--241.

\bibitem{b18}
A.~Gu, K.~Goel, and C.~Re, ``Efficiently modeling long sequences with structured state spaces,'' in \emph{International Conference on Learning Representations}.

\bibitem{b19}
M.-H. Guo, Z.-N. Liu, T.-J. Mu, and S.-M. Hu, ``Beyond self-attention: External attention using two linear layers for visual tasks,'' \emph{IEEE Transactions on Pattern Analysis and Machine Intelligence}, vol.~45, no.~5, pp. 5436--5447, 2022.

\bibitem{b20}
F.~Yang, J.~Luan, M.~Meng, and Y.~Wang, ``Improving bilingual tts using language and phonology embedding with embedding strength modulator,'' in \emph{Proc. INTERSPEECH}, 2023, pp. 5531--5535.

\bibitem{b21}
W.~Hang, W.~Feng, R.~Du, S.~Liang, Y.~Chen, Q.~Wang, and X.~Liu, ``Cross-subject eeg signal recognition using deep domain adaptation network,'' \emph{IEEE Access}, vol.~7, pp. 128\,273--128\,282, 2019.

\bibitem{b22}
K.~Goel, A.~Gu, C.~Donahue, and C.~R{\'e}, ``It’s raw! audio generation with state-space models,'' in \emph{International Conference on Machine Learning}.\hskip 1em plus 0.5em minus 0.4em\relax PMLR, 2022, pp. 7616--7633.

\bibitem{b23}
A.~Gulati, J.~Qin, C.-C. Chiu, N.~Parmar, Y.~Zhang, J.~Yu, W.~Han, S.~Wang, Z.~Zhang, Y.~Wu \emph{et~al.}, ``Conformer: Convolution-augmented transformer for speech recognition,'' 2020.

\bibitem{b24}
H.~Qu, L.~Ning, R.~An, W.~Fan, T.~Derr, X.~Xu, and Q.~Li, ``A survey of mamba,'' \emph{arXiv preprint arXiv:2408.01129}, 2024.

\bibitem{b25}
H.~Shan, A.~Gu, Z.~Meng, W.~Wang, K.~Choromanski, and T.~Sainath, ``Augmenting conformers with structured state-space sequence models for online speech recognition,'' in \emph{ICASSP 2024-2024 IEEE International Conference on Acoustics, Speech and Signal Processing (ICASSP)}.\hskip 1em plus 0.5em minus 0.4em\relax IEEE, 2024, pp. 12\,221--12\,225.

\bibitem{b26}
O.~Lieber, B.~Lenz, H.~Bata, G.~Cohen, J.~Osin, I.~Dalmedigos, E.~Safahi, S.~Meirom, Y.~Belinkov, S.~Shalev-Shwartz \emph{et~al.}, ``Jamba: A hybrid transformer-mamba language model,'' \emph{arXiv preprint arXiv:2403.19887}, 2024.

\bibitem{b27}
P.~Glorioso, Q.~Anthony, Y.~Tokpanov, J.~Whittington, J.~Pilault, A.~Ibrahim, and B.~Millidge, ``Zamba: A compact 7b ssm hybrid model,'' \emph{arXiv preprint arXiv:2405.16712}, 2024.





\bibitem{b28}
L.~Bollens, B.~Accou, M.~Gillis, W.~Verheijen, T.~Francart \emph{et~al.}, ``Sparrkulee: A speech-evoked auditory response repository of the ku leuven, containing eeg of 85 participants,'' 2023.



\end{thebibliography}
\end{document}